\documentclass[twocolumn,twoside,slac_two,unsortedaddress]{revtex4}
\usepackage{graphicx}
\usepackage{fancyhdr}
\usepackage{txfonts}
\usepackage{mathrsfs}

\begin{document}

\title{Pulsar Emission Geometry and Accelerating Field Strength}

\author{Megan E. DeCesar$^{1,2}$, Alice K. Harding$^2$, M. Coleman
  Miller$^1$, Ioannis Contopoulos$^3$, Constantinos
  Kalapotharakos$^{2,3}$, Damien Parent$^4$}
\affiliation{$^1$Department of Astronomy, University of Maryland, College
  Park, MD 20742, USA}
\affiliation{$^2$Astrophysics Science Division, NASA Goddard Space Flight Center, Greenbelt, MD 20771, USA}
\affiliation{$^3$Research Center for Astronomy, Academy of Athens, Athens 11527, Greece}
\affiliation{$^4$Center for Earth Observing and Space Research,
  College of Science, George Mason University, Fairfax, VA 22030,
  resident at  Naval Research Laboratory, Washington, D.C. 20375, USA}

\begin{abstract}
The high-quality {\it Fermi} LAT observations of gamma-ray pulsars
have opened a new window to understanding the generation mechanisms of
high-energy emission from these systems. The high statistics allow for
careful modeling of the light curve features as well as for phase
resolved spectral modeling. We modeled the LAT light curves of the
Vela and CTA 1 pulsars with simulated high-energy light curves
generated from geometrical representations of the outer gap and slot
gap emission models, within the vacuum retarded dipole and force-free
fields. A Markov Chain Monte Carlo maximum likelihood method was used
to explore the phase space of the magnetic inclination angle, viewing angle,
maximum emission radius, and gap width. We also used the measured
spectral cutoff energies to estimate the accelerating parallel
electric field dependence on radius, under the assumptions that the
high-energy emission is dominated by curvature radiation and the
geometry (radius of emission and minimum radius of curvature of the
magnetic field lines) is determined by the best fitting light curves
for each model. We find that light curves from the vacuum field more
closely match the observed light curves and multiwavelength
constraints, and that the calculated parallel electric field can place additional constraints on the emission geometry.

\end{abstract}

\maketitle

\thispagestyle{fancy}

\section{INTRODUCTION}

The pulsar emission mechanism is not well understood. Magnetospheric
particle acceleration is likely responsible for the observed emission,
but the emission geometry is unknown. One can gain some insight by
comparing light curves derived from geometrical emission models with
observed pulsar light curves. Observations of pulsars by the
\textit{Fermi Gamma-ray Space Telescope} Large Area Telescope
(LAT)~\cite{Atwood2009} have shown that the high-energy emission
likely originates in the outer magnetosphere~\cite{PSRCat}. We
simulated high-energy light curves from geometrical versions of two
standard high-altitude emission models, the outer gap
(OG)~\cite{Romani_OG} and slot gap (SG)~\cite{Muslimov_SG}, and from a
third, modified SG model with azimuthal asymmetry in emissivity due to
a naturally occurring offset dipole (aSG)~\cite{Harding_aSG}. These
models were considered within two field geometries, the vacuum
retarded dipole (VRD) and force-free (FF)~\cite{Contopoulos_FF}
fields. We compared the resulting light curves with the LAT light
curves of the Vela pulsar and PSR J0007+7303, the CTA 1 pulsar, to
constrain the systems' geometries, and calculated the model-dependent
magnitude of the accelerating electric field $E_{||}$ in the VRD
field.

\section{LIGHT CURVE MODELING}

To model the LAT light curves, we first simulated pulsar light curves
from geometrical representations of the SG, aSG, and OG emission zones
within the VRD and FF fields, following the simulation method
of~\cite{Dyks2004}. The $B$ field defined in the observer's frame is
transformed to the co-rotating frame (CF)~\cite{Bai_FF} and photons
are emitted tangent to {\bf B} in the CF prior to calculation of the
aberration. We assume constant emissivity along the field lines in the
CF. The azimuthal asymmetry in the polar cap (PC) angle for the aSG model
is calculated for each inclination angle $\alpha$ as
in~\cite{Harding_aSG} (see also~\cite{Harding_theseproc}). For a given
$\alpha$, gap width $w$ (in units of open volume coordinates
$r_{\mathrm{ovc}}$, as in~\cite{Dyks2004}), and maximum emission
altitude $r$, the code outputs the dimensionless emission intensity
and the minimum and maximum radii of curvature, $\rho_{\mathrm{min}}$
and $\rho_{\mathrm{max}}$, emission radii $r_{\mathrm{min}}$ and
$r_{\mathrm{max}}$, and local field magnitude $|B|_{\mathrm{min}}$ and
$|B|_{\mathrm{max}}$, at all observer angles $\zeta$ and rotation
phases $\phi$.

We simulated light curves for a fiducial rotation period of 0.1 s on a
4-dimensional grid of $\alpha$, $\zeta$, $w$, and $r$.
Our simulation resolutions are $1^{\circ}$ in $\alpha$ for the VRD
field and $15^{\circ}$ for the FF magnetosphere; $1^{\circ}$ in
$\zeta$; $0.01\,r_{\mathrm{ovc}}$ in $0\,r_{\mathrm{ovc}} \leq w \leq
0.3\,r_{\mathrm{ovc}}$; and $0.1\,R_{\mathrm{lc}}$ in
$0.7\,R_{\mathrm{lc}} \leq r \leq
2.0\,R_{\mathrm{lc}}$, where $R_{\mathrm{lc}} = c/\Omega$ is the light
cylinder radius.  Emission is allowed out to a cylindrical radius
$r_{\mathrm{cyl}} = 0.98\,R_{\mathrm{lc}}$ for the OG model and
$r_{\mathrm{cyl}} = 0.95\,R_{\mathrm{lc}}$ for the SG models.

\begin{figure*}[t]
\includegraphics[width=\textwidth]{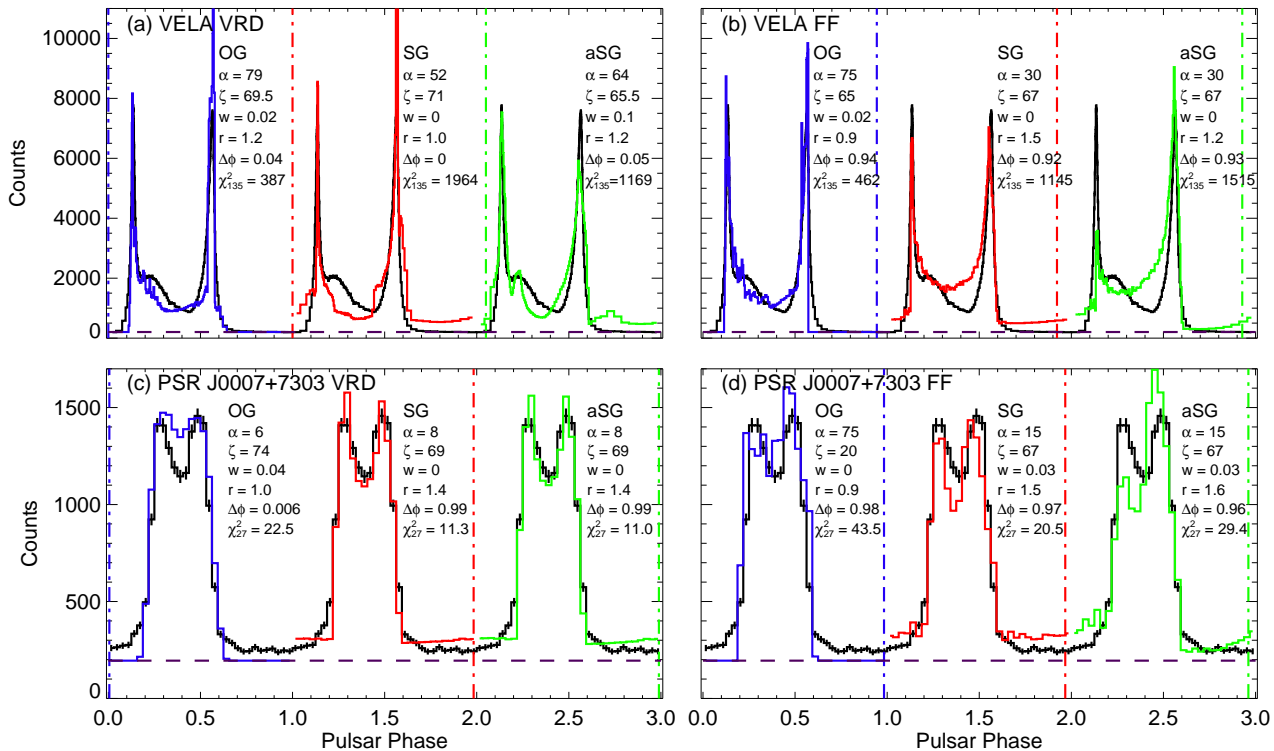}
\caption{Light curve modeling results for Vela and PSR J0007+7303. The
  absolute best fit parameters are given next to each model;
  uncertainties in individual parameters vary, but are typically of
  order $\sim 10$\%. In all panels, the light curve is shown in black.
  Blue curves show the best fit OG light curve, red the best fit SG
  light curve, and green the best aSG light curve. Vertical dot-dashed
  lines show phase zero of the model light curve of corresponding
  color; the phase of the line is the same as the listed $\Delta
  \phi$, and is the phase of emission coming from the closest magnetic
  pole. The horizontal purple dashed lines show the background count
  level. Panel {\it (a)} shows results for the Vela pulsar with the
  VRD field, {\it (b)} Vela with the FF field, {\it (c)} PSR
  J0007+7303 with the VRD field, and {\it (d)} PSR J0007+7303 with the
  FF field.
  \label{lightcurves} }
\end{figure*}

LAT light curves were constructed from photons in an angular radius
$\theta < \mathrm{max}[1.6-3\mathrm{log}_{10}(E), 1.3]$ from the
pulsar~\cite{Vela_2}. The PSR J0007+7303 light curve has 32
fixed-width bins, and was taken from~\cite{CTA1_2}. The Vela light
curve has 140 fixed-count bins of $\sim 3000$ photons each. The
background of PSR J0007+7303, 195 counts/bin, was found using the
\textit{Fermi} tool gtsrcprob as in~\cite{CTA1_2}. The emission in the off-peak
above the background level was assumed to be magnetospheric in origin
(but see~\cite{CTA1_2} for details). The Vela background was found to
be 204 counts/bin using the off-peak (phases 0.8-1, where no
magnetospheric emission was detected in our spectral fits) counts in
the energy-dependent PSF of the pulsar.

We used a Markov Chain Monte Carlo (MCMC) maximum likelihood
routine~\cite{Verde2003} to search the parameter space for the
combination of ($\alpha$, $\zeta$, $w$, $r$, $\Delta \phi$) that best
reproduced the LAT light curves. The fifth parameter, $\Delta \phi$,
is the amount by which a model light curve must shift in phase in
order to best match the LAT light curve. The MCMC begins at a random
point in parameter space, calculates the likelihood $L$, and moves in
a random direction to a new point in space to calculate $L$ again. If
$L/L_{\mathrm{former}} > 1$ or $> \mathrm{rand}[0, 1)$, $L$ is saved
in a chain; the routine runs until the chain contains the
user-specified number of steps. Many chains were run to explore the
whole parameter space. We used Wilks' theorem, $\Delta\,$ln$\,L =
-\Delta \chi^2/2$, to calculate $L$ at each point in parameter space.
To perform a fit, we subtracted the background level from the LAT
light curve, re-binned the model to match the data, and normalized the
model to the total counts in the LAT light curve. For each fit in the
vacuum case, we ran 200 chains with 20 steps each to adequately sample
the parameter space. For the FF fits, we ran 20 chains of 20 steps
each for each $\alpha$. Multi-wavelength constraints on $\alpha$ and
$\zeta$ were considered after fitting. Our fit results are shown in
Figure~\ref{lightcurves}. Each plot shows three instances of the LAT
light curve with the best OG, SG, and aSG light curve superposed. The
zero phases of each model, corresponding to the nearest magnetic pole,
are given by the vertical lines, and the background levels with
horizontal lines.

We find that for the {\it (a)} VRD and {\it (b)} FF fields, the OG model (blue)
statistically fits the Vela light curve better than either SG model.
The SG (red) produces too much off-peak emission, leading to high
$\chi^2$ values. The aSG (green) reduces the background significantly,
leading to a much better fit. The aSG also qualitatively fits the peak
emission well, as its main peaks are the correct approximate height
and an inner peak is present. All three fits have $\zeta$ close to the
value determined from the X-ray torus geometry [10], $\zeta \sim
64^{\circ}$. $\Delta \phi$ is consistent with the observed phase lag
between the radio and $\gamma$-ray peaks for the VRD models; however,
for the FF case, $\Delta \phi$ is too large.

For the pulsar in CTA 1, the SG models fit much better than the OG in
both field geometries. There is no constraint on $\Delta \phi$ due to
the lack of a radio detection. The VRD geometry in {\it (c)} produces much
better fits than the FF in {\it (d)}; there is little difference between the SG
and aSG due to the small $\alpha$. A larger FF $\alpha$ leads to a
larger PC offset, which lowers the first peak. The large difference
between $\alpha$ and $\zeta$ is consistent with the pulsar being
radio-quiet due to geometry--the radio beam would not cross our line
of sight for such a large $|\alpha - \zeta|$.

\section{CALCULATION OF $E_{||}$}

The $\gamma$-ray spectra of pulsars are well fit by an exponentially
cut-off power law,

\begin{equation}
\frac{dN}{dE} = N_0 \left ( \frac{E}{E_0} \right )^{-\Gamma}
\exp{\left [ - \left ( \frac{E}{E_{\mathrm{c}}} \right ) ^b \right ] }
\, \mathrm{ph} \,\, \mathrm{cm}^{-2} \, \mathrm{s}^{-1} \, \mathrm{MeV}^{-1}
\end{equation}

\noindent where $N_0$ is the differential flux, $E_0$ the energy
scale, $\Gamma$ the power law index, and $E_{\mathrm{c}}$ the cutoff
energy; $b = 1$ results in a simple exponential cutoff, while $b < 1$
gives a sub-exponential cutoff and $b > 1$ a super-exponential cutoff.
For phase averaged spectra, $b < 1$ due to blending of
$E_{\mathrm{c}}$ as it varies with phase (e.g.~\cite{Vela_2}), while
$b$ is consistent with (and is fixed to) 1 in individual phase bins.
 
In current models of pulsar emission, at energies above $\sim
100\,$MeV the emission is dominated by curvature radiation. Particles
reach the radiation reaction limit at Lorentz factors
$\gamma_{\mathrm{CR}} \sim 10^7$. In this limit, the curvature
radiation cutoff energy $E_{\mathrm{CR}}$ is related to $E_{||}$ and
$\rho_{\mathrm{c}}$ by

\begin{equation}
E_{\mathrm{CR}} = \frac{3}{2} \frac{\lambdabar}{\rho_{\mathrm{c}}} \gamma^3_{\mathrm{CR}} = 0.32 \lambda_{\mathrm{c}} \left ( \frac{E_{||}}{e} \right )^{3/4} \rho_{\mathrm{c}}^{1/2}
\end{equation}

\begin{figure}[ht]
\includegraphics[width=0.5\textwidth]{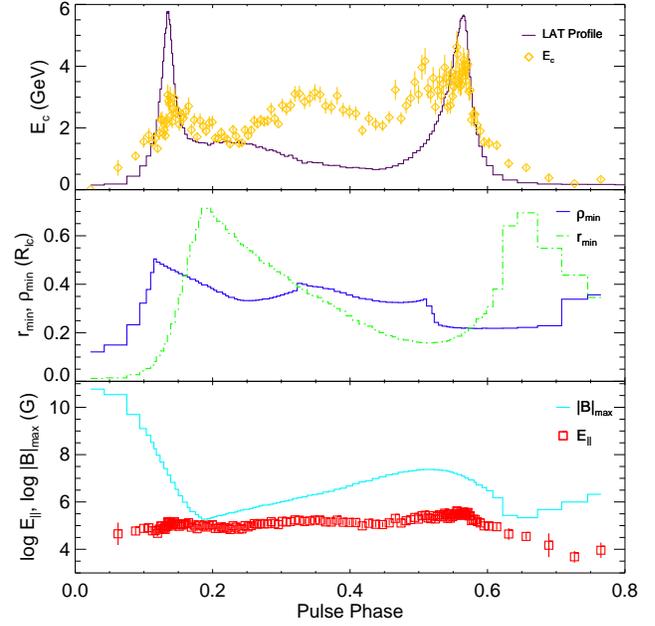}
\caption{Example of simulated and measured values described in \S3.
  {\it (a)} The LAT light curve of Vela (purple) and measured
  $E_{\mathrm{c}}$ with phase (yellow diamonds).  {\it (b)} Simulated
  $\rho_{\mathrm{min}}$ (solid blue) and $r_{\mathrm{min}}$ (dashed
  green) for the best VRD SG fit parameters. {\it (c)} Simulated
  $|B|_{\mathrm{max}}$ (solid cyan) and calculated $E_{||}$ (red
  squares) for the best VRD SG fit parameters.  \label{3panels} }
\end{figure}

\noindent Assuming all emission with $E > 100\,$MeV is due to pure
curvature radiation and that the VRD is the true $B$ field structure,
we calculated $E_{||}$ in each light curve phase bin. We used the
simulated minimum radii of curvature ($\rho_{\mathrm{c}} =
\rho_{\mathrm{min}}$) from the best fit VRD light curves of \S3 and
the measured cutoff energies ($E_{\mathrm{CR}} = E_{\mathrm{c}}$) in
each phase bin. The cutoff energies are given in~\cite{CTA1_2} for PSR
J0007+7303. We updated the Vela 0.1--100 GeV phase resolved spectral
results with 30 months of LAT data, following the method
of~\cite{Vela_2} with 3000 pulsed counts per bin, and used our
measured cutoff energies for the Vela $E_{||}$ calculation. As an
example, Figure~\ref{3panels} shows the measured $E_{\mathrm{c}}$, the
simulated $\rho_{\mathrm{min}}$, $r_{\mathrm{min}}$, and
$|B|_{\mathrm{max}}$, and the calculated $E_{||}$ for the Vela pulsar
peak emission (phases $0 \leq \phi \leq 0.8$), using the best fit
parameters from the VRD SG geometry.

\begin{figure*}[t]
\centering
\includegraphics{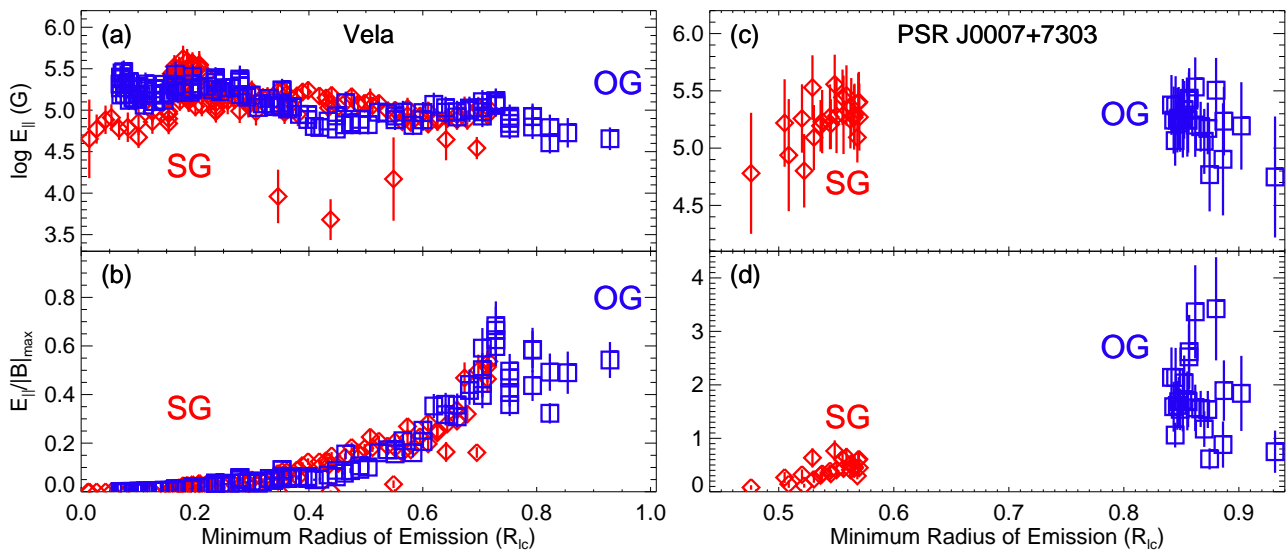}
\caption{$E_{||}$ {\it (a)} and the ratio of $E_{||}/|B|_{\mathrm{max}}$
  {\it (b)} for the OG (blue squares) and SG (red diamonds) models
  that best fit the Vela LAT light curve. {\it (c)} Same as {\it (a)}
  for PSR J0007+7303. {\it (d)} Same as {\it (b)} for PSR J0007+7303.
  \label{eparallel} }
\end{figure*}

We explore how the parallel electric field varies with emission
altitude. We have calculated $E_{||}$ in each phase bin for the best
fit vacuum OG and SG model parameters.
Because the value of $|B|_{\mathrm{max}}$ corresponds to
$r_{\mathrm{min}}$, we have plotted $E_{||}$ and
$E_{||}/|B|_{\mathrm{max}}$ with minimum emission radius for the OG and
SG models in Figure~\ref{eparallel}.

For both emission models, Vela has overall a $\sim\,$constant or
gradually varying $E_{||}$ with altitude (panel $(a)$ of
Figure~\ref{eparallel}), which is expected (e.g.~\cite{Vela_2}).
In panel $(b)$, the value of $E_{||}$ is compared with
$|B|_\mathrm{max}$. As expected for an $E$ field induced by $B$, the
ratio $E_{||}/|B|_{\mathrm{max}} < 1$ for all $r_{\mathrm{min}}$ out to
the light cylinder radius (near and beyond $1\,R_{\mathrm{lc}}$, the
vector components of {\bf B} are less certain and are not included in
this calculation).

For PSR J0007+7303, the magnitude of $E_{||}$ with $r_{\mathrm{min}}$
is consistent with a constant (panel $(c)$), and its values are
similar to those calculated for Vela. Note that the geometrical
parameters obtained from the light curve fits are very different from
those of Vela, leading to a different range of $r_{\mathrm{min}}$ for
PSR J0007+7303. The ratio $E_{||}/|B|_{\mathrm{max}} < 1$ for the best
fit SG model, but it is $> 1$ for the best fit parameters of the OG
model. There are instances in, for example, a non-ideal
magnetosphere~\cite{Kalapotharakos2011} where $E_{||}$ may be $> |B|$.
In the case of the vacuum field, in which there are no currents, the
only source of $E_{||}$ is induction by $B$, and therefore $E_{||}$
cannot be larger than $|B|_{\mathrm{max}}$. In this particular case,
then, we find that
the combination of the OG model and VRD field we have used does not
approximate the physical environment of the pulsar magnetosphere
and/or the geometry of the emission zone.

\section{CONCLUSIONS}

We have evaluated the geometries of the slot gap and outer gap
emission models, and the vacuum retarded dipole and force-free magnetic
field solutions, by comparing the simulated light curves with the LAT
light curves of Vela and PSR J0007+7303 and finding the geometrical
model parameters that best reproduce the data in each case. In
general, the OG has no off-peak emission (this is largely responsible
for the OG fitting Vela the best and PSR J0007+7303 the worst), while
the SG models do a better job of reproducing wing emission but
over-predict the off-peak emission. Introducing azimuthal asymmetry of
the PC angle in the aSG model leads to a reduction in the off-peak
emission level, improving upon the SG model light curve fits within
the VRD field. In the FF field, however, aSG light curves tend to have
a much reduced first peak, leading to significantly worse light curve
fits.

The first $\gamma$-ray peak occurs at later phase in FF light curves,
so values of $\Delta \phi$ are larger in the FF magnetosphere than in
the VRD field. Physically, $\Delta \phi$ cannot be larger than the
phase lag between the radio and first $\gamma$-ray peak unless the
radio beam model is highly contrived. The FF field requires a $\Delta
\phi$ larger than this phase lag for Vela. This suggests that the true
pulsar magnetosphere may be significantly different from the FF
magnetosphere.

For Vela, all fits within the VRD field have $\zeta$ close to the
expected value of $64^{\circ}$, and all have reasonable values of
$|\alpha - \zeta|$ such that the radio emission is observable.
Interestingly, the aSG model has $\zeta$ closest to $64^{\circ}$, and
while its $\chi^2$ is poor, it is an improvement over the SG and
qualitatively reproduces well the major features (two main peaks and
inner peak) of the pulsed emission. The FF fits also get close to the
correct $\zeta$ value, but only the OG has an acceptable $|\alpha -
\zeta|$, while the best fit SG models are consistent with a
radio-quiet pulsar. Both field structures lead to large $|\alpha -
\zeta|$ for PSR J0007+7303, consistent with the lack of detected radio
emission.

We calculated the model-dependent $E_{||}$ for the OG and SG
geometries, and found that it is constant or slowly varying with
emission radius. Comparing $E_{||}$ with $|B|_{\mathrm{max}}$ leads to
the interesting result that for PSR J0007+7303, $E_{||} >
|B|_{\mathrm{max}}$ for the OG parameters that best fit the LAT light
curve. This is not consistent with the VRD where $|E| < |B|$, and thus
disfavors the OG model in the VRD geometry at the location in
parameter space where the best light curve fit is found. The
calculation of $E_{||}$ can therefore be used as a diagnostic of the
model magnetosphere and emission geometry, in addition to the light
curve fitting.

We cannot rule in favor of the SG or OG from these light curve fits.
As the models are purely geometrical, we expect to reproduce only
dominant light curve features, and our statistical fits are poor.
The vacuum field produces better light curve fits
than the force-free field. However, we note that the resolution in
$\alpha$ of our FF models is much lower than in the VRD, so further
modeling with higher resolution is needed to confirm this
result. We have demonstrated that light curve modeling leads to
constraints on the geometry of individual systems, and that the phase
lag is an important diagnostic in comparing magnetic field structures.
The model-dependent calculation of $E_{||}$, and comparison with the
model $|B|_{\mathrm{max}}$, can additionally be used to constrain both
the magnetosphere and emission geometry.

\begin{acknowledgments}
The \textit{Fermi} LAT Collaboration acknowledges the institutes
supporting the LAT development, operation, and data analysis. MED
acknowledges support from CRESST and the \textit{Fermi} Guest Investigator
Program.
\end{acknowledgments}


\end{document}